# DETECTING AND MEASURING RESPIRATORY EVENTS IN HORSES DURING EXERCISE WITH A MICROPHONE: DEEP LEARNING VS. STANDARD SIGNAL PROCESSING

**Jeanne I.M. Parmentier**[0000−0002−3141−8106]
Utrecht University (Utrecht, NL)
University of Twente (Enschede, NL)
Inertia Technology B.V. (Enschede, NL)
`j.i.m.parmentier@uu.nl`

**Rhana M. Aarts**[0009−0000−9058−465X]
Utrecht University (Utrecht, NL)
`r.m.aarts@uu.nl`

**Elin Hernlund**[0000−0002−5769−3958]
Swedish University of Agricultural Sciences (Uppsala, SE)
`elin.hernlund@slu.se`

**Marie Rhodin**[0000−0003−0575−2765]
Swedish University of Agricultural Sciences (Uppsala, SE)
`marie.rhodin@slu.se`

**Berend Jan van der Zwaag**[0000−0003−0884−5334]
University of Twente (Enschede, NL)
`b.j.vanderzwaag@utwente.nl`

September 18, 2025

## ABSTRACT

Monitoring respiration parameters such as respiratory rate could be beneficial to understand the impact of training on equine health and performance and ultimately improve equine welfare. In this work, we compare deep learning-based methods to an adapted signal processing method to automatically detect cyclic respiratory events and extract the dynamic respiratory rate from microphone recordings during high intensity exercise in Standardbred trotters. Our deep learning models are able to detect exhalation sounds (median F1 score of 0.94) in noisy microphone signals and show promising results on unlabelled signals at lower exercising intensity, where the exhalation sounds are less recognisable. Temporal convolutional networks were better at detecting exhalation events and estimating dynamic respiratory rates (median F1: 0.94, Mean Absolute Error (MAE) ± Confidence Intervals (CI): 1.44±1.04 bpm, Limits Of Agreements (LOA): 0.63±7.06 bpm) than long short-term memory networks (median F1: 0.90, MAE±CI: 3.11±1.58 bpm) and signal processing methods (MAE±CI: 2.36±1.11 bpm). This work is the first to automatically detect equine respiratory sounds and automatically compute dynamic respiratory rates in exercising horses. In the future, our models will be validated on lower exercising intensity sounds and different microphone placements will be evaluated in order to find the best combination for regular monitoring.

**K**eywords Artificial intelligence · Automated detection · Horse · Respiration · Respiratory rate · Microphone · Exercise physiology.

# 1 Introduction

Respiration parameters, including breathing patterns, respiratory rate (RR), inhalation-exhalation ratio, respiratory sounds and locomotor-respiratory coupling represent critical physiological aspects in exercising horses. Quantitative and qualitative analyses of these parameters during exercise offer significant potential for understanding performance mechanisms in horses [1], detecting respiratory anomalies and ultimately improving equine welfare.

Horses are obligate nose breathers meaning that they cannot switch to oronasal breathing like humans. The equine upper respiratory tract (URT) includes the nostrils, nasal passages, pharynx, and larynx and is essential for moving air in and out during respiration, with inspiratory and expiratory airway flows reaching 65-75 L/s and 60-80 L/s respectively during exercising [2]. In comparison, human literature reported maximum nasal airflows of about 3.5 L/s during the inspiratory phase and 4.8 L/s during the expiratory phase [3]. Resulting peak tracheal pressures in horses were reported as -18 mmHg to -37 mmHg during inhalation and 6 mmHg to 20 mmHg during exhalation for normal horses during maximal exercise [2]. Any obstruction can become performance limiting and also a relevant welfare problem as horses may experience breathlessness [4].

Upper respiratory tract disorders are common in horses and may be associated with abnormal respiratory sounds during exercise [5]. For example, noises like high-frequency "whistling" or "roaring" during inspiration are described for horses with laryngeal dysfunction [6, 7] while low-frequency expiratory noises are described for horses with dorsal displacement of the soft palate [7]. Other described respiration abnormalities are tachypnoea (abnormally high respiratory rate), coughing, frequent swallowing [8], breathe holding, or abnormal breathing patterns [5].

The current gold standard for URT disorder diagnosis is via an upper airway endoscopy, either at rest or during exercise. This technique provides medical images of the URT yet requires veterinary intervention and interpretation. Moreover, endoscopic analysis is a (minimally) invasive procedure with risks of discomfort for the animal and of damaging the mucosal membranes during the insertion of the endoscope through the nasal cavity. Specialized equine face masks were also developed to quantify airflow and other respiratory parameters during exercise [9, 10, 11] but they restrict ventilation [12], are often impractical and can therefore be considered invasive too. Digital stethoscopes with radio-transmission were developed [13] and used in different studies to characterise respiratory sounds [14], locomotor-respiratory coupling [15] or response to exercise [16]. The quality of the recorded sounds using stethoscopes is highly dependent on the placement and the absence of friction with the hair or skin during locomotion.

In the past decades, several studies have explored the use of microphones to study respiratory sounds and characterised sounds and patterns relative to pathologies, showing promising perspectives to detect upper-airway disorders in horses [17, 18, 19]. These successful first steps motivate the further use and exploration of microphone data to monitor equine respiration during training. Most studies with microphones were conducted on treadmills with Thoroughbred (gallop races) or Standardbred (trot races) horses. While being extremely valuable for making a diagnosis, treadmill measurements and evaluations are not representative of the field conditions [20], especially not for harness trotters which are trained and raced with a sulky. Few studies implemented microphone measurements during field trials, but their analyses of the respiratory rate were limited by the segmentation of respiratory sounds. These were manually counted to extract an average respiratory rate [1, 21], overlooking fluctuations over time. These fluctuations can be especially interesting in trotting animals for which, contrary to cantering or galloping animals, the respiratory cycle is not always coupled to the locomotor cycle [22, 15].

In the equine field and as far as the authors know, only a single recent explorative work has shown that microphone data can be automatically labelled with exhalation events [23] (F1 scores above 85%), but this study had looked at only one deep learning model type and had not yet explored automatic computation of relevant physiological indicators. The human field is a length ahead as several studies were published in which microphone data were used to monitor various aspects of respiration. Most of these are focused on the detection of specific sound types in lung, tracheal or nasal sound recordings in a controlled environment [24], or during low intensity tasks where the respiration sounds are usually relatively quiet and subsequent respiratory rate is low [25, 26].

In recent years, some works combined the advances in smartphone technology and in (w)earables or small portable microphones to monitor respiratory sounds and extract respiratory rates during higher intensity exercises such as running or general exercising, using either simple signal processing methods [27, 28] or exploiting deep learning models such as recurrent networks and convolutional networks [29] for more advanced and complex tasks. The strength of these models lies in their ability to either learn temporal characteristics of the signals or automatically extract local features relevant for the given task and apply learnt concepts on new, unseen data, making them good candidates for respiratory sounds detection. These models can be used to process audio features, spectrograms or even raw signals, enabling the further exploration of the potential of the recorded sounds.

This study aims at automatically (1) detecting cyclic respiratory events in recordings obtained with a simple off-the-shelf omni-directional stereo microphone attached to the nose of harness trotters and (2) computing respiratory rates using detected events. To this end, we trained two types of sequence-to-sequence deep learning classifiers with manually labelled signals of different sources (microphone channels) and sampling frequencies recorded at high-speed trot. From the outputs of our models, we computed dynamic respiratory rates. In parallel, we also adapted a simple signal processing method published in the human literature [27] to dynamically estimate the respiratory rates from our microphone signals. Finally, we apply our models and the signal processing method to unlabelled audio recordings obtained at lower speed with less respiratory intensity.

## 2    Materials and Methods

### 2.1    Data collection

Fifteen Standardbred trotters in full training were included in this study (5 mares, 5 geldings, 5 stallions, age: 3.0±1.1 years old). Each horse was equipped with an omni-directional stereo microphone (ECM-LV1, Sony, 44100 Hz), and voice recorder (ICD-Px470, Sony). The microphone was taped with adhesive foam bandage (Animal Polster, Snögg®) between the two nostrils, with one channel in the middle (channel 1) and one channel closer to the left nostril (channel 2). The voice recorder was fixed to the cheek piece of the bridle and cables were secured with tape (Figure 1). Horses were also equipped with 9 inertial measurement unit sensors recording 3D accelerations and 3D rotation speeds at 200 Hz (ProMove-mini, Inertia Technology B.V., Enschede, The Netherlands). One of the sensors was also equipped with a Global Navigation Satellite System (GNSS) chip which was used to record speed (1 Hz). The other sensors were not used for this study.

All horses underwent a standardised exercise test for harness trotters on the same oval track (sand surface with particle size: 0-4 mm), with incrementally increasing trotting speeds (for the protocol, see [30, 31] and Supplementary Table S1). Shortly, each horse underwent an adapted standard exercise test during which they had to incrementally increase their speed. The last segment of high-speed trot, where horses were running 500 or 800 m (37.9±1.5 km/h), was used to train the different models in this study. Respiration sounds were more distinguishable (louder with higher signal amplitudes) at higher compared to lower speeds, thus enabling manual labelling of the respiration events. Moreover, high-speed trot is a condition under which the respiratory system of the trotters is under pressure, further motivating this choice.

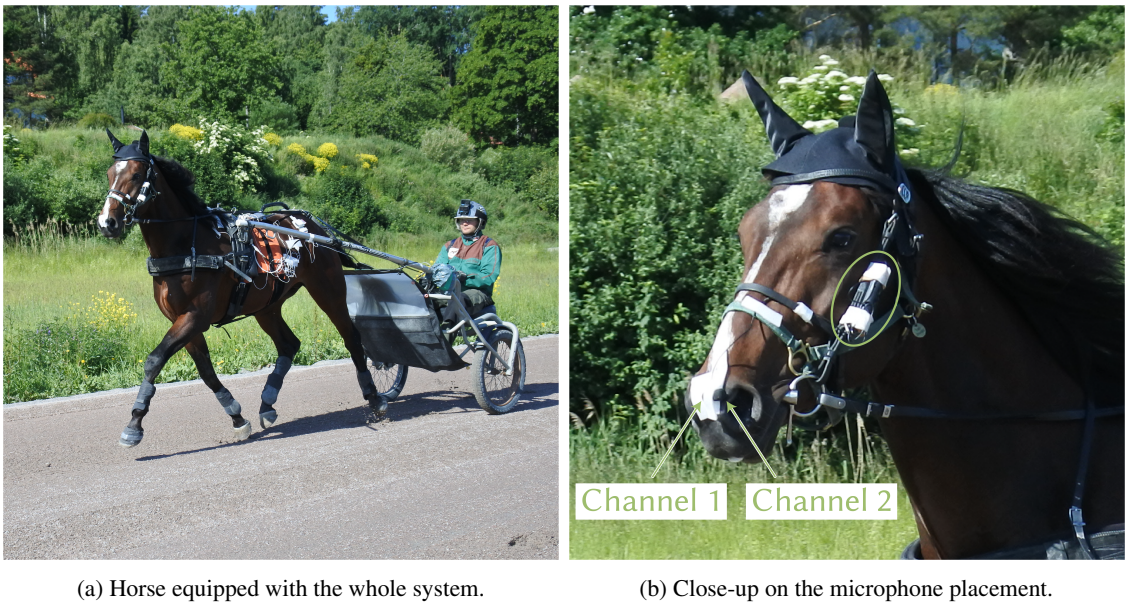

(a) Horse equipped with the whole system.                    (b) Close-up on the microphone placement.

Figure 1: (a) Horse equipped with data collection equipment and performing the standard exercise test and (b) close-up image of the microphone showing channel 1 (left arrow), channel 2 (right arrow) and voice recorder (ellipse) setup. Copyright: Rhana M. Aarts, Utrecht University.

## 2.2 Data processing

Data labelling was conducted in Audacity® 3.4.2. [32], other data processing and models training were conducted in MATLAB R2024a (MathWorks, Natick, Massachusetts, USA).

### 2.2.1 Data segmentation

Data were segmented in speed bouts as defined earlier, using measurement notes of the timestamps. A brief description of average speed for each segment is available in supplementary Table S1.

### 2.2.2 Data labelling

First, the audio recordings from segments of high-speed trot were displayed in Audacity® with both channels in stereo mode. Their respective Mel spectrograms (Scale 1-20000 Hz, Hann window of 2048 samples, zero padding factor 2, gain of 20 dB and range of 80 dB) were also displayed. One author (J.I.M.P.) identified and labelled the exhalation sounds and patterns in the microphone data, using a slower playback speed if needed. Segments in which the horse was neighing were considered as "noise" and not further analysed. Labelled events were checked together with another author (R.M.A.). In case of doubts regarding an event, consensus was reached between the users by listening and visualising the surrounding events. Subsequently, *.txt* files with event timestamps were exported to MATLAB for further analysis. For each audio file, an associated label time series was created with 0 when there was no exhalation event and 1 when there was an exhalation event (see Figure 2 for an illustration).

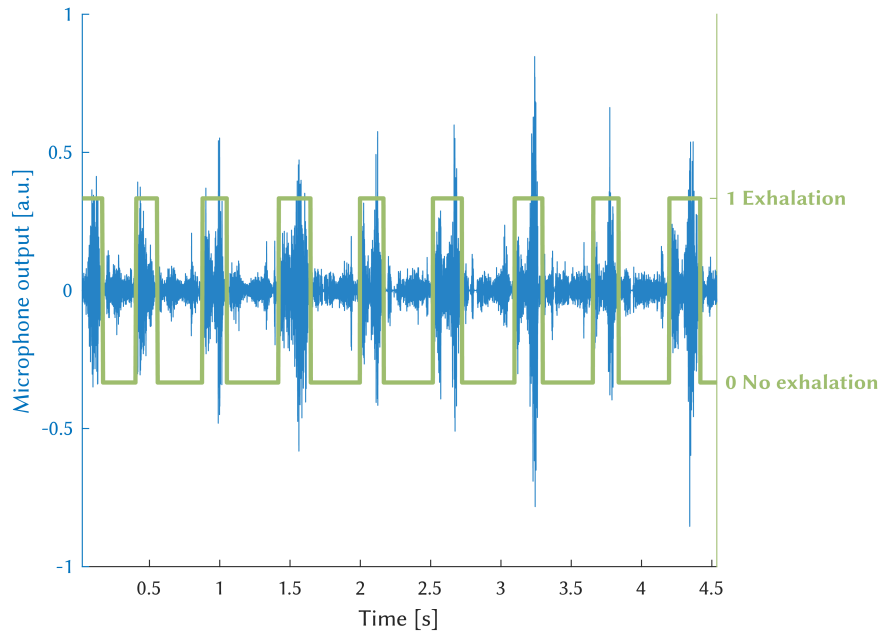

Figure 2: Example of microphone signal of one horse (channel 2, blue) and associated manual labelling (green).

### 2.2.3 Downsampling

Because near-real-time applications require lower data rates, we also evaluated the effect of sampling frequency on the automated detection of exhalation events. We downsampled the signals to 4410 Hz, 2100 Hz, 1050 Hz and 490 Hz. The MATLAB function *downsample* without offset was used with downsampling factors of 10, 21, 42 and 90 respectively. An example of downsampled signals is shown in Figure 3.

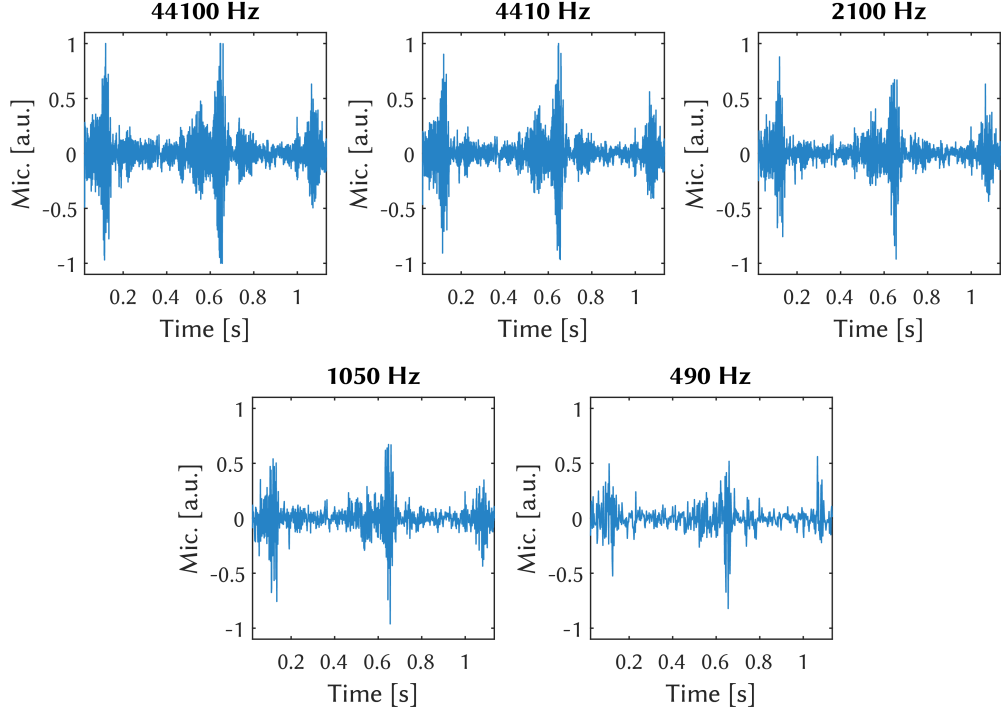

Figure 3: Example of downsampled signals for channel 1 for different sampling frequencies, compared to the initial sampling frequency (44100 Hz).

## 2.3 Respiratory events detection

### 2.3.1 Architecture of the models

For this work, we aimed at sequence-to-sequence classification (from microphone signals to "No Exhalation" (0) and "Exhalation" (1)) and compared two deep-learning model architectures: Long Short-Term Memory Network (LSTM) [33] and Temporal Convolutional Network (TCN) [34]. These two models were specifically chosen as they were already proved to be particularly well suited for event detection in biomedical signals [35, 23]. With the goal of reducing the complexity of the models while obtaining good detections, we also compared the performance of the models when trained with both stacked microphone channels as an input, or only channel 1 or only channel 2.

For both LSTM and TCN, the data were first passed into a sequence input layer with a rescale-symmetric normalisation and the networks ended with a full connected layer, a softmax layer and a classification layer, with a cross-entropy loss function. Different depths of LSTM and TCN were evaluated, as described below. All models and depths were trained with the adam optimiser [36], a MiniBatch size of 8, an initial learning rate of 0.001, a validation patience of 10 epochs and a maximum of 100 training epochs. The best model training iteration, based on the lowest validation loss, was kept for testing.

*LSTM*
Three depths of LSTM were tested in this work: 1, 2 and 4 blocks. Each block included a bilateral LSTM layer with 50 hidden units and a dropout layer as presented in Figure 4.

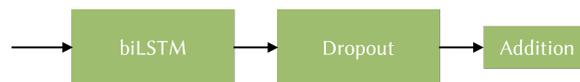

Figure 4: Building block of an LSTM model.

*TCN*
Three depths of TCN were tested in this work: 4, 8 and 12 blocks. Each block is constituted of a 1D convolution network, a normalisation layer, a spatial dropout layer, another convolution network, another normalisation layer, a reLU layer and a spatial dropout layer as presented in Figure 5.

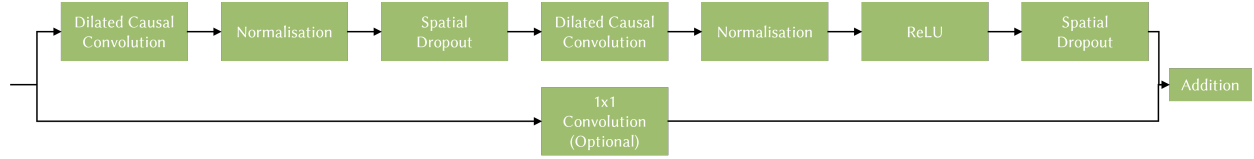

Figure 5: Building block of a TCN model.

### 2.3.2 Training and performance evaluation

Across all horses, 947 s of high-speed trot audio data were extracted. A total of 36 TCN detectors ((3 input types) x (4 sampling frequencies) x (3 building block depths)) and 36 LSTM detectors ((3 input types) x (4 sampling frequencies) x (3 building block depths)) were trained and tested (see Table 1). To train and evaluate our detectors, we used the Leave-One-Out (LOO) method, meaning that each model architecture was trained and evaluated 15 times with a different train-validation-test set. Each horse was kept unseen once for the test set, one horse was randomly selected to form the validation set and the remaining horses were used for training. For each testing iteration, a confusion matrix was built and the precision, recall and subsequently F1 score were computed. If an F1 score could not be computed (i.e., recall and precision equal to 0), a score of 0 was attributed. The best models were the ones with higher median F1 score and narrower interquartile range.

Table 1: Summary of the input data (microphone channel and sampling frequency) used to train the Long Short-Term Memory (LSTM) and Temporal Convolutional Network (TCN) models of different depths.

| Input | channel 1, channel 2 or both channels |
|---|---|
| **Sampling frequency** | 490, 1050, 2100 or 4410 Hz |
| **Model** | LSTM or TCN |
| **Depth LSTM** | 1 block, 2 blocks or 4 blocks |
| **Depth TCN** | 4 blocks, 8 blocks or 12 blocks |

### 2.3.3 Application to unlabelled data

Unlabelled downsampled data (4410 Hz and 2100 Hz) from segments of lower intensity (Trot 1 in Supplementary Table S1, average speed of 25.11±2.33 km/h) were used as input in the TCN and LSTM models. Time needed to compute the outputs of the models was quantified using the *tic toc* function of MATLAB and was expressed as a percentage of the duration of the input data.

## 2.4 Respiratory rate estimation

For this section, the audio recordings were further segmented in subwindows of 10 seconds with a 50% overlap sliding window. These subwindows of data were used to compute respiratory cycle durations and subsequently dynamic average respiratory rates as described below.

### 2.4.1 Reference

The manually labelled events were used as the reference method. For each subwindow $w$ of 10 seconds, the duration of the $N$ respiratory cycles ($T^w(n)$) was computed (eq. 1) and used to calculate the average respiratory rate $RR_{Reference}(w)$ (eq. 2) such as:

$$T^w(n) = t_{event}(n+1) - t_{event}(n) \tag{1}$$

$$RR_{Reference}(w) = \frac{60}{N}\sum_{n=1}^{N}\frac{1}{T_n^w} \tag{2}$$

An example is shown in Figure 6, top panel, where the green line represents the reference signal and the black diamonds the reference respiratory events.

### 2.4.2   Deep learning method

The output of each DL model (LSTM, TCN) was post-processed for each subwindow of 10 seconds before RR estimation, in order to remove exhalation events that were detected erroneously. The following steps were followed:

1. Computation of the moving average (0.1 sec) to smooth the output (MvAvg)

2. Thresholding of the MvAvg: $\geq 0.1$ is Exhalation, $<0.1$ is No Exhalation

3. Computation of the moving variance (0.1 sec) of the absolute microphone signal used (if both channels, the average of channel 1 and channel 2 is used) (MvVar)

4. For each exhalation detected, if the MvVar is below 0.8 then the Exhalation event is discarded.

The respiratory events were defined as the beginning of the detected exhalation events. For each subwindow, the envelope of the microphone signal was extracted, detrended using the *detrend* function in MATLAB and peaks were detected with a minimum distance of 0.30 seconds [37]. If no peak was present within a respiratory event, it was considered as erroneous detection and was discarded. The duration of the respiratory events and the average RR over each window was computed as described in equations 1 and 2 respectively. Events with a distance below 0.30 seconds [37] were discarded.

An example is shown in Figure 6, middle panel, where the green line represents a post-processed DL output signal and the black diamonds the obtained respiratory events.

### 2.4.3   Signal processing method

A signal processing (SP) method was also applied to the microphone signals (channel 1 or channel 2), based on the method described in [27] (differences are indicated by *). For the whole segment of high-speed data, the following steps were followed:

1. Removal of the mean of the microphone signal

2. Filtering of the signal:
   (a) for the signals downsampled at 2100 Hz and 4410 Hz: bandpass filter with cutting frequencies of [200, 800] Hz
   (b) for the signals downsampled at 1050 Hz and 490 Hz: high-pass filter with cutting frequency of 200 Hz

3. Computation of the absolute of the Hilbert transform of the signal, and subsequent smoothing with a moving average filter (0.1 second)

4. Down or upsampling of the resulting output to 1000 Hz

5. Bandpass filter with cutting frequencies of [0.01, 3]* Hz

Then, for each subwindow of 10 seconds**, the cyclic respiration events were obtained by:

1. Detrending of the windowed signal, using the *detrend* function in MATLAB***

2. Finding peaks with a minimum distance of 0.30**** seconds and a minimum height of 2% of the maximum peak height in the subwindow $w$ of interest

3. Computation of the respiratory events duration, where an event is a previously found peak, and subsequent computation of the average RR (equations 1 and 2 resp.)

in [27]:

* the last bandpass filter uses cutting frequencies of [0.01, 2] which was too low for the equine respiration in this work;

** windows of 30 seconds were used to find the cyclic respiration events and compute an RR value. As our horses breathe faster than what was described for humans, dynamic RR changes can occur at a higher rate than every 30 seconds and we therefore decided to compute our RR values on shorter windows of time;

*** this step is not described;

**** the minimum distance between two peaks is 0.70 seconds which was too high for the equine respiration during high intensity exercises [37].

An example is shown in Figure 6 bottom panel, where the green line represents the final SP output signal and the black diamonds the obtained respiratory events as previously described.

### 2.4.4 Reference, DL and SP comparison

To compare the DL and SP methods to the reference RR, the mean absolute error (MAE) of each testing iteration $n$ was computed such as:

$$MAE_{Method}(n) = \frac{1}{W} \sum_{w=1}^{W} \left| RR_{Reference}(w) - RR_{Method}(w) \right| \qquad (3)$$

where $W$ is the number of subwindows in the $n^{\text{th}}$ testing iteration.

Subsequently the overall average, standard deviation (s.d.) and confidence intervals (CI) of the MAE for each method were computed such as:

$$mean\ MAE_{Method} = \frac{1}{N} \sum_{n=1}^{N} MAE_{Method}(n) \qquad (4)$$

$$s.d.\ MAE_{Method} = \sqrt{\frac{1}{N-1} \sum_{n=1}^{N} (MAE_{Method}(n) - mean\ MAE_{Method})^2} \qquad (5)$$

$$CI\ MAE_{Method} = 1.96 \times \frac{s.d.\ MAE_{Method}}{\sqrt{N}} \qquad (6)$$

where $N$ is the number of testing iterations, i.e. one for each test horse.

Additionally, means of differences (MOD) and 95% limits of agreements (LOA) between the reference and the DL and SP methods were computed and used to obtain Bland-Altman plots [38]. MOD and LOA were computed such as:

$$MOD_{Method} = \frac{1}{M} \sum_{m=1}^{M} RR_{Reference}(m) - RR_{Method}(m) \qquad (7)$$

where $M$ is the number of RRs extracted for each method.

$$LOA_{Method} = \pm 1.96 \times s.d.(RR_{Reference} - RR_{Method}) \qquad (8)$$

### 2.4.5 Application to unlabelled data

The outputs from the DL models and the unlabelled data described in section 2.3.3 were post-processed following the steps in section 2.4.2 and average respiratory rates were computed for each subsegment. Similarly, the raw microphone signals of unlabelled data were computed with the SP method as described in section 2.4.3 to obtain RR values.

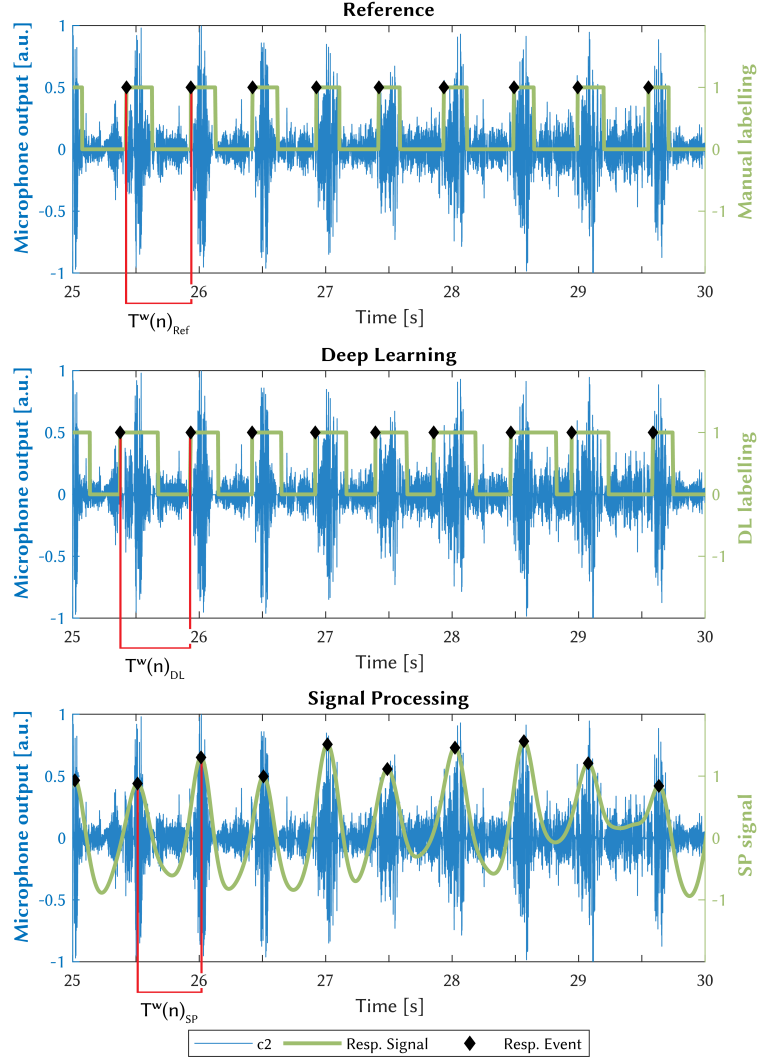

Figure 6: Comparison of the respiratory event obtention for the reference method (manual labelling of the Exhalation events, top panel), the deep learning method (automated labelling of the Exhalation events, middle panel) and the signal processing method (bottom panel). This example is obtained with the channel 2 microphone output (blue lines), downsampled at 4410 Hz. The signals used for respiratory events obtention are shown with the green lines, the respiratory events used to compute the duration of the respiratory cycles ($T^w(n)$) are highlighted by black diamonds. *Ref*: Reference, *DL*: deep learning, *SP*: signal processing, *c2*: channel 2, *Resp. Signal*: respiratory signal, *Resp. Event*: Respiratory event.

## 3   Results

### 3.1   Noise and respiratory sounds

The audio files did not only reveal respiratory sounds. Other types of sound and noise like jangling of the tack, hoof beats on the ground and talking of the drivers, together with environmental sounds such as the chirping of birds and the wind could be identified. Neighing of the horses was easily detected and showed to saturate the microphone sensors.

Of the respiratory sounds, the exhalations were the most recognizable and thus reliable to label at faster speeds (i.e., "normal trot" and "fast trot" compared to lower speeds as during the warming-up or cooling-down). Regardless of the speed, inhalation sounds were harder to detect, especially in shallow breathing horses. One of the proposed reasons is that environmental noises drown the inhalation sounds at lower speeds, since less air is transported during breathing, when compared to higher speeds which require more effort and thus more strenuous breathing. Interestingly, different respiration patterns were heard between horses in terms of frequency, consistency and type of sound. Some horses appeared to be very constant in their breathing pattern (inhalation-exhalation-inhalation), whereas others skipped a breath, possibly due to swallowing, or appeared to show two small in- or exhalation bouts instead of one long breath (inhalation-inhalation-exhalation). The types of sounds varied from clean respiration with no additional or unexpected noises, to respiration with "floppy nostrils" (otherwise described as "high blower" in [14]) or respiration with roaring sounds. Examples of different respiration sounds and patterns are available in the Supplementary Materials.

### 3.2 Respiratory events detection

#### 3.2.1 Labelled data

The detector performances are shown in Figure 7 and Table 2.

The best performing models were the TCN with 8 blocks, trained with both channels sampled at 1050, 4410 or 2100 Hz (median F1 score: 0.944, 0.941 and 0.940 and IQR: 0.022, 0.038 and 0.026 respectively), followed by TCN models of 12 blocks and 8 blocks trained with channel 2 sampled at 4410 Hz (median F1 score: 0.939 and IQR: 0.028 and 0.046 respectively). Similar results are obtained with TCN 8 and 12 blocks, trained with both channels sampled at 4410 and 1050 Hz respectively. Overall, LSTM models showed lower median F1 scores and larger IQR values than TCN, especially when trained with channel 1.

Better results were obtained with both channels or channel 2, compared to channel 1. The best detection results of channel 1 were obtained with signals sampled at 2100 Hz and the TCN 8 blocks (F1 median 0.913 IQR 0.071). LSTMs trained with channel 1 data showed larger IQR values, regardless of the depth or the sampling frequency.

For the LSTMs of 2 and 4 blocks trained with channel 1 signals, increasing the sampling frequency decreased the detection performance. Long Short-Term Memory models with a depth of 2 blocks and trained with both channels showed a slight increase of performance with higher sampling frequencies, especially with narrower IQR values (F1 median 0.842 IQR 0.121 at 490 Hz and F1 median 0.881 IQR 0.051 at 4410 Hz). Temporal Convolution Networks were less sensitive to the downsampled frequency when trained with channel 2 and both channels. When trained with channel 1, increasing the sampling frequency slightly increased the median F1 score and reduced the IQR values.

Shallower LSTMs with a depth of 1 block showed a worse performance compared to deeper models but doubling the depth from 2 to 4 blocks did not radically increase the exhalation detection performance. Similarly, increasing the depth of the TCN slightly increased the performance when trained with data from channel 1 but there were no remarkable differences for models trained with channel 2 or both channels. The optimum combination was the TCN 8 blocks trained with both channels, with sampling frequencies above 490 Hz.

Table 2: Results of the best 10 (before double-line) and worst 2 exhalation detectors (after double-line), based on the median and interquartile range (IQR) values of the F1 scores. *TCN*: Temporal Convolutional Network, *LSTM*: Long Short-Term Memory, *Fs*: sampling frequency, *c1*: channel 1, *c2*: channel 2, *both*: both channels, *IQR*: interquartile range.

| Rank | Model | Depth | Input | Fs (Hz) | F1 score Median | F1 score IQR |
|------|-------|-------|-------|---------|-----------------|--------------|
| 1 | TCN | 8 | both | 1050 | 0.944 | 0.022 |
| 2 | TCN | 8 | both | 4410 | 0.941 | 0.038 |
| 3 | TCN | 8 | both | 2100 | 0.94 | 0.026 |
| 4 | TCN | 12 | c2 | 4410 | 0.939 | 0.028 |
| 5 | TCN | 8 | c2 | 4410 | 0.939 | 0.046 |
| 6 | TCN | 12 | both | 4410 | 0.937 | 0.034 |
| 7 | TCN | 12 | c2 | 1050 | 0.937 | 0.034 |
| 8 | TCN | 12 | both | 2100 | 0.937 | 0.034 |
| 9 | TCN | 8 | c2 | 1050 | 0.937 | 0.037 |
| 10 | TCN | 12 | both | 1050 | 0.936 | 0.036 |
| | LSTM | 2 | c1 | 4410 | 0.774 | 0.147 |
| | LSTM | 4 | c1 | 4410 | 0.764 | 0.353 |

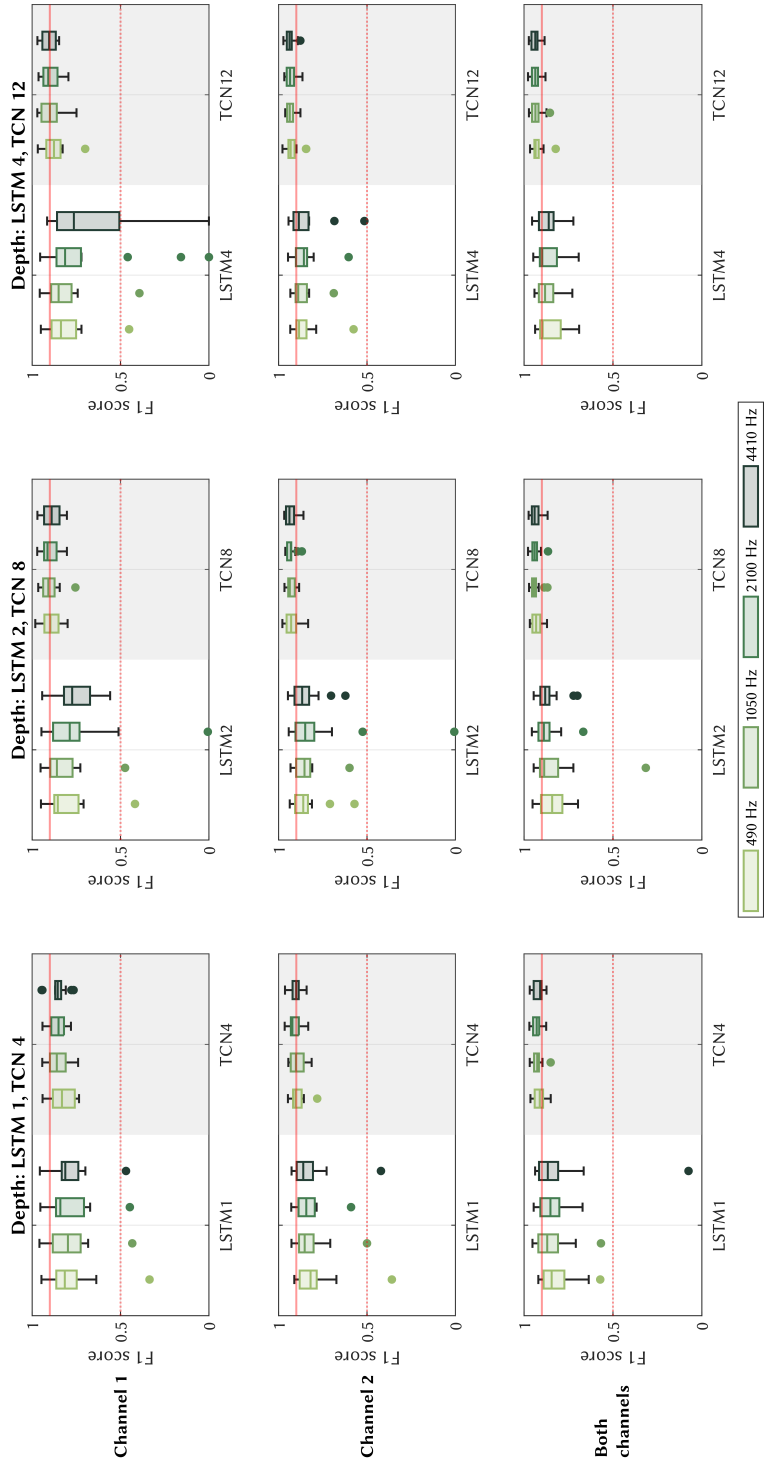

Figure 7: Performance of the different Long Short-Term Memory (LSTM) and Temporal Convolutional Network (TCN) detectors expressed as F1 score for each sampling frequency (490, 1050, 2100 and 4410 Hz). The columns represent the depth of the LSTM or TCN (1 block/4 blocks are the shallowest detectors, 4 blocks/12 blocks are the deepest respectively). Each row corresponds to the input microphone data (channel 1, channel 2, both channels). Dots represent classification outliers, full red lines represent the F1 score of 0.9 (target), dotted red lines represent the F1 score of 0.5 (random detector).

### 3.2.2 Unlabelled data

Due to the lack of reference labels, F1 scores were not computed for the unlabelled data. However, visual comparison of the outputs of the models and the signals showed that the models were able to detect exhalation events. These events were used to compute RR values, further presented in section 3.3. A couple of raw output promising examples are presented in the Supplementary Figures S2a and S2b while a couple of unfavourable examples are presented in the Supplementary Figures S3a and S3b.

The computation durations needed for each DL model per input and sampling frequencies are shown in Figure 8. Temporal Convolutional Networks were much faster than LSTM models regardless of the input data or the sampling frequency used. For example, TCN of 4 blocks required 3.89% of the segment duration to label the data of both channels sampled at 4410 Hz while LSTM required 5.58% of the segment duration. Downsampling the signals from 4410 to 2100 Hz generally decreased the computation duration of the models by almost half, regardless of the model type or the input data. There were no differences in computation durations between the TCN models trained with same depth using different input of data of the same sampling frequency. However, the computation durations of LSTMs were more sensitive to the input data used. Durations were always shorter when using both channels with LSTM of depth 1 block or when using channel 1 with LSTM of depth 2 and 4 blocks.

Model sizes ranged from 90 KB (LSTM 1 block) to 2005 KB (TCN 12 blocks). There were no differences in model size between models trained with data sampled at 4410 or 2100 Hz and a difference of 1.2-1.6 KB between the models trained with one or both channels.

## 3.3 Respiratory rate estimation

### 3.3.1 Labelled data

Respiratory rate results are shown in Table 3 and Figures 11, 10a and 10b and in Figure 9 for each horse. Deep learning models that were not able to detect exhalation events for at least one testing iteration were disregarded for this section as no RRs could be computed reliably (n = 1/36 LSTM models). All SP and TCN models were able to compute RRs, regardless of the input data or the downsampling rate used.

Overall, TCN models performed better with smaller MAE and CI values, narrower LOA (Table 3), and good correlations and agreements with the reference RR values obtained with the manual labels (Figures 10a and 10b). Using lower sampling frequencies slightly decreased the estimation performance when using both channels or channel 2 (TCN 8 blocks, both channels: MAE±CI 1.44±1.04 bpm and 1.76±1.30 bpm, LOA: 7.06 bpm and 9.27 bpm, at 4410 and 2100 Hz respectively) but better estimations were obtained at 490 Hz compared to 1050 Hz for TCN 8 blocks, with both channels or channel 2 (Figure 9). The best TCN model presented a MOD of 0.63, meaning a slight overall underestimation of the RR values. This is especially observable for H5 and H9 in Figure 9.

In comparison, LSTM models outputs were not as good to compute RR values and presented larger LOA. Best LSTM results were obtained with a depth of 4 blocks and with both channels sampled at 4410 Hz (MAE±CI: 3.52±3.11 bpm, LOA: 14.03 bpm), or sampled at 2100 Hz with 1 or 4 blocks (MAE±CI: 4.61±5.70 bpm and 4.90±5.91 bpm, LOA: 21.08 bpm and 18.23 bpm respectively).

Overall, SP methods adapted from [27] using channel 2 performed well with MAE and CI values below 4 bpm and LOA below 13 bpm. The best SP results were obtained with channel 2 sampled at 4410 Hz (Table 3); these displayed good correlations and agreements with the RR obtained with the manual labels. Using lower sampling frequencies decreased the RR estimation performance for the SP method regardless of the input data used (Figure 11).

While using channel 1 led globally to the worst results for the DL models with MAE values above 10 bpm and LOA above 35 bpm, TCN models of 12 blocks and SP methods were still able to estimate RR values using channel 1 sampled at 4410Hz (MAE±CI: 2.47±1.11 bpm and 3.27±1.89bpm, LOA: 11.35 bpm and 14.28 bpm).

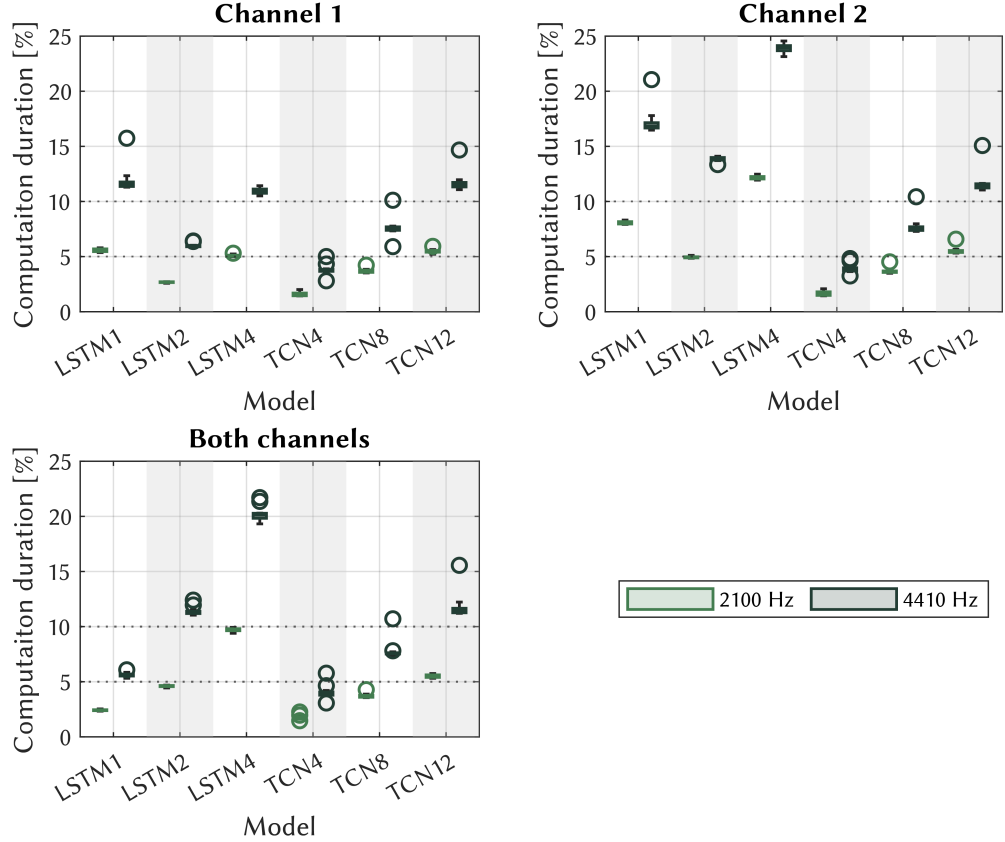

Figure 8: Boxplots of the exhalation events labelling computation duration expressed as percentage of the total segment duration, per model type and depth, sampling frequency and input data. Dotted lines serve as indicators for the 5% and 10% marks. *LSTM*: Long Short-Term Memory, *TCN*: Temporal Convolutional Network.

Table 3: Results of the best 10 respiratory rate estimators, based on the MAE and CI range. *Fs*: sampling frequency, *MAE*: mean absolute error, *CI*: confidence intervals, *MOD*: means of differences, *LOA*: limits of agreement, *bpm*: breaths per minute, *TCN*: Temporal Convolutional Network, *SP*: signal processing, *c1*: channel 1, *c2*: channel 2, *both*: both channels.

| Rank | Model | Depth | Input | Fs (Hz) | MAE±CI (bpm) | MOD (bpm) | LOA (bpm) |
|------|-------|-------|-------|---------|--------------|-----------|-----------|
| 1 | TCN | 8 | both | 4410 | 1.44±1.04 | 0.63 | 7.06 |
| 2 | TCN | 12 | both | 4410 | 1.67±1.19 | 0.48 | 7.7 |
| 3 | TCN | 8 | both | 2100 | 1.76±1.30 | 0.36 | 9.27 |
| 4 | TCN | 12 | both | 2100 | 2.25±1.31 | -0.18 | 9.48 |
| 5 | TCN | 12 | c2 | 4410 | 2.42±1.69 | -0.99 | 12.5 |
| 6 | TCN | 8 | c2 | 4410 | 2.43±1.33 | -0.56 | 10.33 |
| 7 | TCN | 12 | c1 | 4410 | 2.47±1.11 | 0.73 | 11.35 |
| 8 | SP | n.a. | c2 | 4410 | 2.48±1.48 | 0.31 | 10.5 |
| 9 | TCN | 12 | c2 | 2100 | 2.67±1.60 | -0.9 | 13.43 |
| 10 | TCN | 8 | both | 490 | 2.64±1.69 | -0.51 | 10.49 |

### 3.3.2 Unlabelled data

Results are shown in Figures 12 and 13, as well as in Table 4. Overall, our average RR values are below the ones reported in Cotrel et al. [1]. In their work, reported speeds were higher (minimum 30km/h) compared to the speeds of the segment of unlabelled data used for in this study (average 25.11 km/h).

With the DL models, average RR values increased depending on the input used. They were lowest when computed with channel 1 and highest when computed with channel 2. This difference can be explained by the fact that the

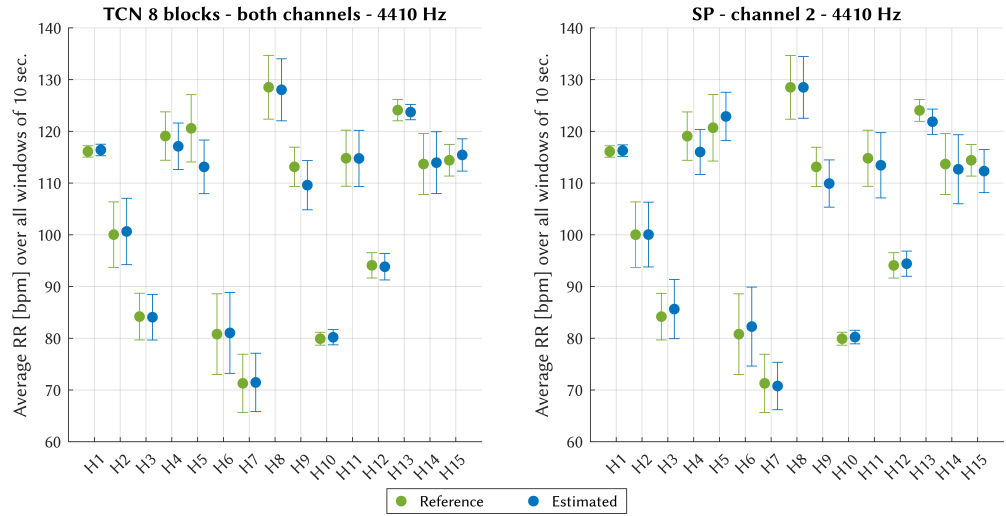

Figure 9: Mean and confidence intervals values of reference and estimated respiratory rates across all extracted subwindows per horse (top panels) obtained with the outputs of TCN 8 blocks, using both channels sampled at 4410 Hz (left) or with the SP method applied to channel 2 sampled at 4410 Hz (right). *RR*: respiratory rate; *TCN*: Temporal Convolutional Network; *LSTM*: Long Short-Term Memory; *SP*: signal processing; *bpm*: breaths per minute.

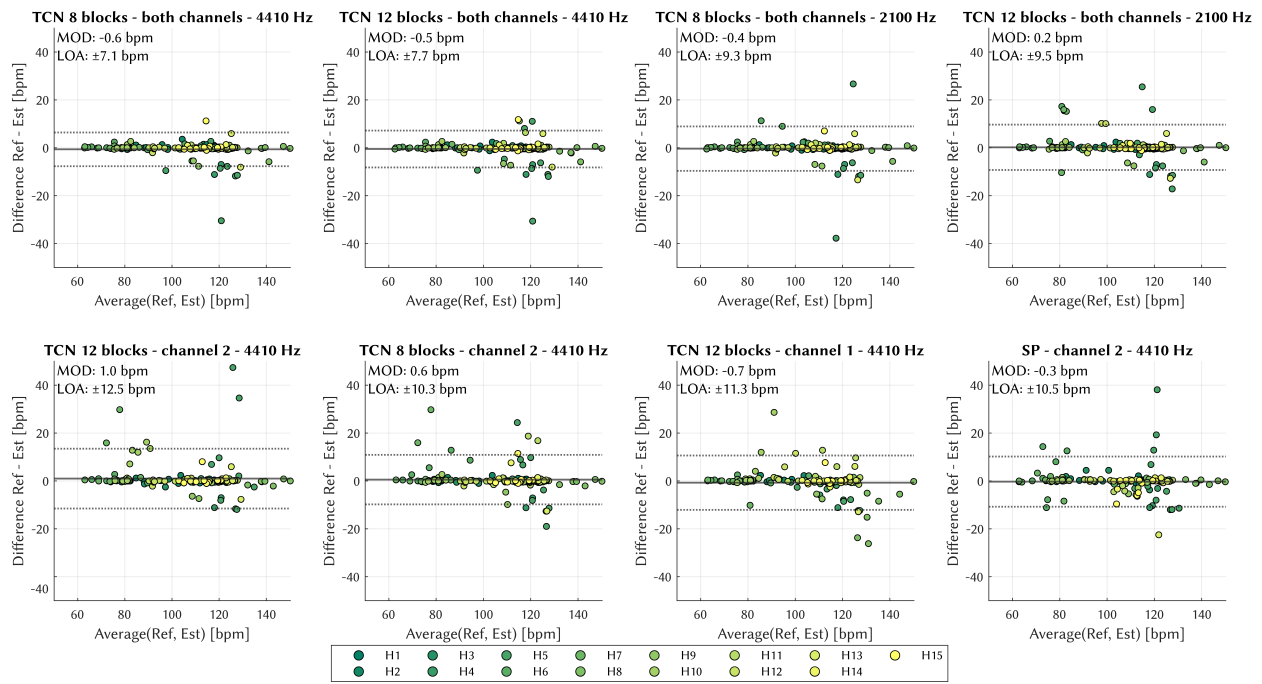

(a) Bland-Altman plots

signal-to-noise ratio was much lower at low-intensity exercising, especially in channel 1, compared to higher-intensity segments used to train the models. Moreover, the DL output post-processing steps used a threshold value applied to the microphone signal(s) of interest. Channel 1 is the channel further away from the nostrils compared to channel 2, and some peaks might have been removed during the post-processing steps, as the amplitude of the respiratory sounds captured by channel 1 may be too low. Likewise, RR obtained with both channels are also lower than those from channel 2 only, as an average of the two signals was used. This is especially visible in Figure 13 for horses H2, H8, H10 or H15, suggesting that the sounds could not be properly detected by our proposed methods with both channels.

All methods detected low RR values for H7 (<50 bpm), regardless of the input data. Upon checking, we manually counted about 21 exhalation sounds for 30 seconds of recording, which led to a respiratory rate of 42 bpm. On the other hand, for H15, we counted about 49 exhalation sounds for 30 seconds of recording leading to a RR of 98 bpm. The SP model estimated this RR value correctly using channel 2 but the DL models underestimated this value.

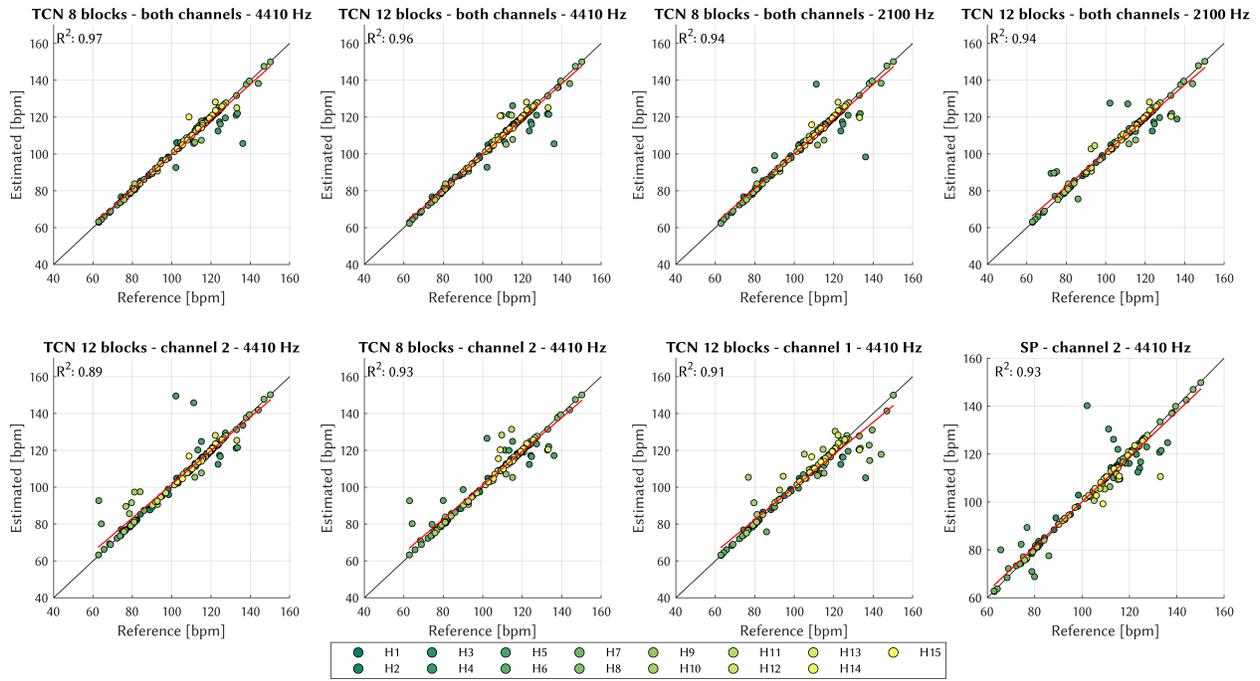

(b) Correlation plots

Figure 10: (a) Bland-Altman plots comparing the respiratory rates obtained with the best 8 methods to those obtained with the reference method (manual selection). Each dot represents the average respiratory rate computed over a subwindow of 10 seconds. Full lines represent the MOD; dotted lines represent MOD±LOA. (b) Corresponding correlation plots. Black lines represent perfect linear relationship between the two methods. Red lines represent the fitted linear regression model between the two methods. $R^2$ values show the fitness of the linear regression model. *SP*: signal processing, *TCN*: Temporal Convolutional Network, *LSTM*: Long Short-Term Memory, *c1*: channel 1, *c2*: channel 2, *Ref*: reference, *Est*: estimated, *MOD*: means of differences, *LOA*: limits of agreement, *bpm*: breaths per minute.

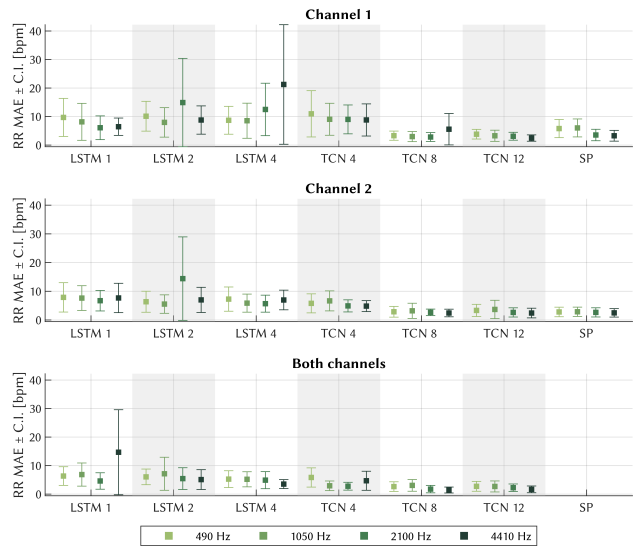

Figure 11: Mean absolute errors and associated confidence intervals for respiratory rates computation per input data, model and model depths. *RR*: respiratory rate, *MAE*: mean absolute error, *CI*: confidence intervals, *LSTM 1, 2, 4*: Long Short-Term Memory network of depth 1, 2 or 4 blocks, *TCN 4, 8, 12*: Temporal Convolutional Network of depth 4, 8 or 12 blocks, *SP*: signal processing.

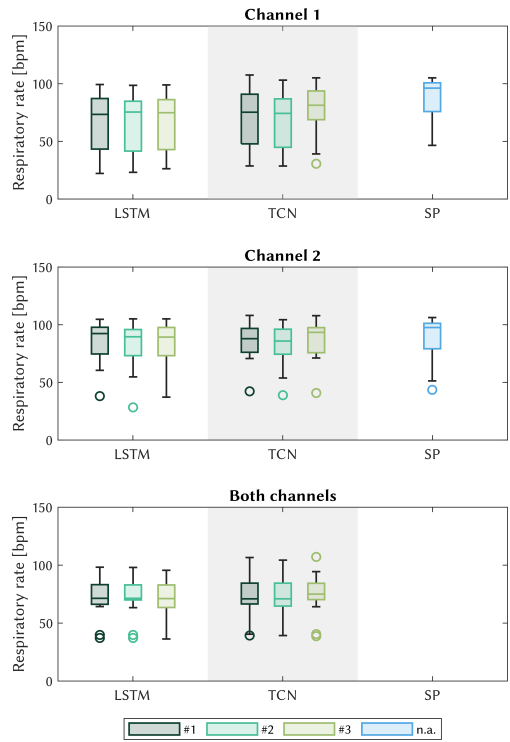

Figure 12: Boxplots of the average respiratory rates computed with channel 1, channel 2 or both channels and different depths of LSTM and TCN models, and with the SP method for each horse over a whole segment of unlabelled data at a lower intensity. *LSTM*: Long Short-Term Memory, *TCN*: Temporal Convolutional Network, *SP*: signal processing, *#1, #2, #3*: LSTM or TCN depth, such as #1: 1 block (LSTM) or 4 blocks (TCN); #2: 2 blocks (LSTM) or 8 blocks (TCN); #3: 4 blocks (LSTM) or 12 blocks (TCN), *bpm*: breaths per minute.

Table 4: Mean, standard deviation and confidence interval values of the respiratory rates obtained on unlabelled data. Results are compared to the maximum and minimum values reported in Cotrel et al. [1]. *RR*: respiratory rate, *s.d.*: standard deviation, *CI*: confidence intervals, *LSTM*: Long Short-Term Memory, *TCN*: Temporal Convolutional Network, *SP*: signal processing, *c1*: channel 1, *c2*: channel 2, *both*: both channels, *bpm*: breaths per minute.

| Model | Depth | Input | Mean RR (bpm) | s.d. RR (bpm) | CI (bpm) |
|---|---|---|---|---|---|
| LSTM | 1 | c1 | 68.05 | 24.54 | 12.42 |
| LSTM | 1 | c2 | 84.1 | 18.19 | 9.21 |
| LSTM | 1 | both | 72.61 | 17.26 | 8.73 |
| LSTM | 2 | c1 | 66.76 | 24.15 | 12.22 |
| LSTM | 2 | c2 | 82.42 | 20.12 | 10.18 |
| LSTM | 2 | both | 72.62 | 16.89 | 8.55 |
| LSTM | 4 | c1 | 68.43 | 24.5 | 12.4 |
| LSTM | 4 | c2 | 83.91 | 18.7 | 9.46 |
| LSTM | 4 | both | 70.45 | 17.79 | 9 |
| TCN | 4 | c1 | 72.23 | 24.95 | 12.63 |
| TCN | 4 | c2 | 85.24 | 16.87 | 8.54 |
| TCN | 4 | both | 73.23 | 17.74 | 8.98 |
| TCN | 8 | c1 | 69.62 | 24.08 | 12.19 |
| TCN | 8 | c2 | 82.75 | 18.41 | 9.32 |
| TCN | 8 | both | 71.98 | 18.44 | 9.33 |
| TCN | 12 | c1 | 76.56 | 22.88 | 11.58 |
| TCN | 12 | c2 | 86.13 | 17.13 | 8.67 |
| TCN | 12 | both | 73.93 | 17.63 | 8.92 |
| SP | n.a. | c1 | 87.43 | 18.09 | 9.15 |
| SP | n.a. | c2 | 88.46 | 19.42 | 9.83 |
| Cotrel et al. (max) | n.a. | n.a. | 112.8 | 19.8 | n.a. |
| Cotrel et al. (min) | n.a. | n.a. | 87.6 | 17.4 | n.a. |

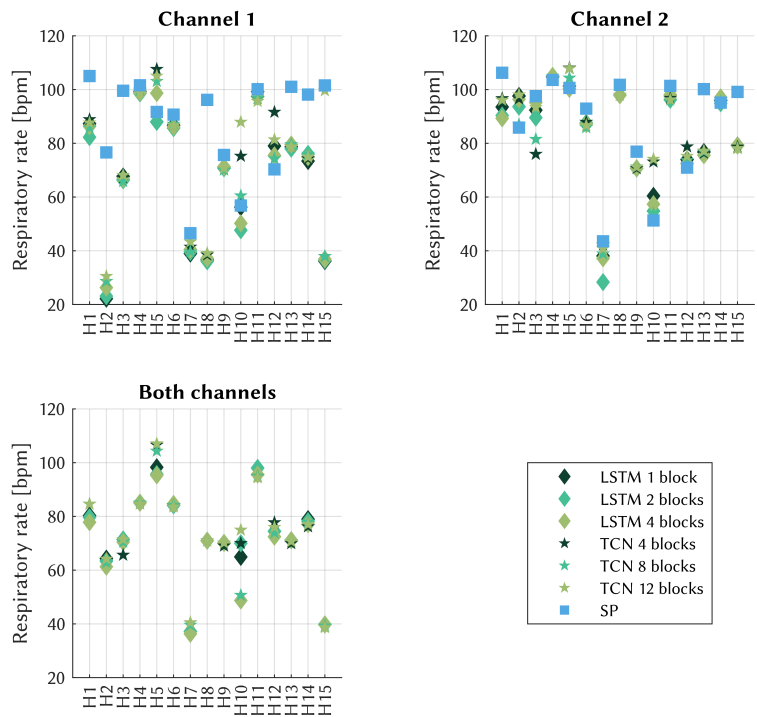

Figure 13: Respiratory rates computed with channel 1, channel 2 or both channels for each horse over a whole segment of unlabelled data at a lower intensity. *bpm*: breaths per minute, *TCN*: Temporal convolutional network, *LSTM*: Long Short-Term Memory, *SP*: signal processing.

## 4 Discussion

This work is the first to demonstrate the feasibility of automated exhalation events detection from downsampled respiratory recordings, and subsequent respiratory rate computation, during high intensity exercising of harness trotters. We showed that using deep learning models, exhalation events could automatically and accurately be detected without extensive signal processing, and that near-real-time respiratory rates could be computed from said events. Finally, we showed that our results outperformed a previously published signal processing method from the human literature that we adapted for our application to equine data.

### 4.1 Respiratory events detection

Detecting respiratory events during exercise is important to monitor changes in breathing patterns and associated physiological parameters. In this work, we focused on the detection of the exhalation events as they were the most prominent sounds in our audio recordings.

Our models trained with both channels performed better than when trained with channel 2 only, and the worst performances were obtained when training with channel 1 only. This is in accordance with our previously published results for TCN models [23]. Channel 2 was the closest to the source of the sound of interest, presented a higher signal-to-noise ratio and would even be able to capture lower amplitude sounds at lower intensity. Moreover, as performances generally increased when including channel 1 to channel 2 in our models, our results suggest that the simultaneous recording of the surrounding noise helps the models to better identify the target sounds by potentially cancelling out the noises present in the signals from channel 2, similarly to standard noise cancellation signal processing methods [39].

Our TCN models performed better than our LSTM models, which is in accordance with empirical results obtained by Bai et al. [34]. Furthermore, TCN models are less computationally costly and the run time to train this kind of model is much lower than the one required by LSTM models, making them interesting candidates for tasks involving large and complex biological signals. They also presented lower computational durations than LSTM models when predicting or classifying input data, which is also useful for (near) real-time applications (Figure 8). However, TCN model files were larger than LSTM ones, which might have been accounted for when designing a system for near real-time respiratory rate monitoring.

Labelling respiratory sounds at lower speed is especially challenging in noisy environments like equine exercising sessions. In the absence of ground truth methodology such as a flowmeter mask, the labelling process relies on the hearing of the respiratory sounds and visualisation of the signals and their spectrograms. Supplementary Figures S1a and S1b show examples of respiratory signals from horses at lower speed and higher speed respectively where the differences in signal amplitude and Mel spectrogram are appreciable. Despite these differences, our models could detect exhalation events in data from a new condition with a lower signal-to-noise ratio and where the respiratory sounds had a lower amplitude compared to higher speed recordings. To confirm these first results, further labelling of our microphone signals will be necessary.

### 4.2 Respiratory rate estimation

Respiratory rate is an important physiological aspect of equine health and performance that has been quantified until now by the manual counting of respiratory events in video recordings (endoscope), radio-stethoscope recordings or audio signals (microphone) of 20.48 seconds and more [1, 21, 14, 16]. In this work, we showed for the first time that equine RR could be automatically estimated from microphone signals in a fine-grained manner in high intensity exercising horses. Our methods accurately detected RR values ranging from 60 to 160 bpm in windows of 10 seconds using standard signal processing methods (MAE±CI: 2.48±1.48 bpm, MOD: 0.31 bpm, LOA: 10.50 bpm) or by using the outputs of our exhalation events detectors (MAE±CI: 1.44±1.04 bpm, MOD: 0.63 bpm, LOA: 7.06 bpm).

In our work, most good performing DL models slightly underestimated the RR values (positive means of differences), which was due to a slightly early detection of the exhalation event, while other DL models slightly overestimated RR values (negative means of differences), which can be due to a confusion between exhalation events and other types of sound from the microphone signals. Extra pre-processing steps to clean the microphone signals or the models' outputs could be helpful to improve detection and consequently respiratory rate estimation. This is also supported by the fact that by adapting a signal processing-based method published in the human literature [27], RR could also accurately be estimated from respiratory sounds at high-speed. Using the SP method, we obtained slightly larger LOA values than the ones at the running highest running speed reported in [27] (10.5 bpm vs 8.93 bpm) but lower MAE and MOD (MAE: 2.48 bpm vs 3.8 bpm and MOD: 0.31 bpm vs -.01 bpm). Considering that our range of respiratory rate is larger (around 60-160 bpm vs around 25-45 bpm), we can conclude that we successfully adapted their method to a highly noisy, dynamic and diverse new application.

The TCN models and the SP method were more robust to the microphone location than the LSTM models. While better results were obtained with channel 2 than with channel 1 with the SP method, the differences were marginal (best results: MAE 2.36 bpm, LOA 10.43 bpm with channel 2, MAE 3.36 bpm and LOA 14.36 bpm with channel 1). The best DL results for channel 1 were obtained with TCN 12 blocks, which performed slightly worse compared to when using channel 2 only (MAE: 2.47 and 2.42 bpm, LOA: 11.35 and 15.50 bpm, respectively). This is encouraging for testing microphone placements further away from the nostrils, to a more practical location which would allow for an easier use.

Regardless of the method employed, better results were obtained at higher sampling frequencies. For a standalone performance monitoring system, lower sampling rates are preferred as they would lead to lower computational costs and lower battery consumption [40], ultimately providing a wearable solution with low maintenance needs for the users. However, for clinical monitoring or to support diagnosis of URT pathologies, sampling frequencies as high as 8-9 kHz are desired as abnormal sounds were occasionally described in the 2-4kHz band in roarers, at around 3.8 kHz in horses with induced laryngeal hemiplegia or in the 1.1-2.7 kHz range for horses with dynamic laryngeal collapse [17, 18, 19]. In our previously published study, we showed that excellent exhalation events detection performances were obtained with signals downsampled to 8820 Hz [23], which could therefore be used for clinical applications in the future.

### 4.3 Limitations

*Amount of data*
Our study is limited by the low number of horses and exercising intensities included in the analysis which could lead to overfitting from the deep learning models, despite the high between- and within-individuals variations in the respiratory sounds and respiratory rates. To take this into account, we implemented safeguarding steps to avoid overfitting in our deep learning models such as dropout layers and early stopping based on validation loss. In the future, adding more labelled data should improve the performance and robustness of our models.

*Optimal models and labels*
While we conducted a hyperparameter search, we focused on a limited range of possible hyperparameters. More optimal DL models might exist outside of the ones defined in this study, especially regarding the LSTM architecture. As LSTM models were more computationally demanding, we did not explore further this architecture in the context of this study. Moreover, only two users labelled the data and the final labels, mainly from one user, were used after consensus. Due to the high sampling rate of the microphone signals and due to the nature of audio signals, identifying the exact beginning and end of the exhalation events was complex and there is uncertainty in our final labels, which can also have an impact on our results.

*Health status*
We did not consider the cardiorespiratory health of the population from this study, which could affect the results of our models and of the signal processing method. For example, studies have shown that horses with dynamic laryngeal collapse present high frequency noise during inhalation [19], which could be detected by our models or by the signal processing method as an event and therefore artificially increase the estimated respiratory rate.

*Microphone placement*
The final limitation is the microphone placement. The microphone was taped to the nose of the horses, which can be unpractical in everyday practice. In the future, we will collect respiratory data with different microphone placements and consequently fine-tune and re-evaluate our models accordingly.

## 5  Conclusion

In conclusion, this study showed that relevant equine respiratory sounds during high intensity exercising can be captured with cheap off-the-shelf microphones, and respiratory events can be automatically detected and labelled by deep learning models from the audio signals. Respiratory rates at high-speed trot can be obtained through signal processing methods or through deep learning by using the previously detected events. At lower speed, both deep learning and signal processing methods seem to be able to estimate respiratory rates, but fine-tuning and validation will be required especially for microphone placements further away from the sound source. These results are promising for different applications such as regular monitoring of health and performance of sport horses but also early detection of abnormal respiratory sounds or monitoring of vocalisation related to the emotional valence [41] during training without the need for more invasive and expensive methods. In the future, other physiological parameters can be estimated such as the inhalation-exhalation ratio to further assess the temporal precision of our models. Additionally, synchronising microphone signals to motion sensor signals will allow for a more holistic health and performance monitoring and will allow further exploration of equine respiratory mechanisms.

**Ethical statement**

**Data availability**

The audio files used in this study are published and publicly available in [42].

**Acknowledgements**

The authors would like to thank Zala Žgank, Ineke Smit, Ebba Zetterberg and Guilherme de Camargo Ferraz for their help during the data collection, as well as the Menhammar Stuteri drivers and grooms. They would also like to thank René van Weeren, Filipe Serra Bragança and Maarten van Steen for their input regarding this manuscript. The data collection for this project was funded through the Varenne Project (EUREKA Eurostars E!114697).

**Declaration of Interest**

J.I.M.P and B.J.vdZ. are part-time employees of Inertia Technology B.V., which company was part and coordinator of the Varenne Project. J.I.M.P. started her work on the project before her employment at Inertia Technology B.V. and B.J.vdZ. did the work for this manuscript as part of his other part-time appointment at the University of Twente. Inertia Technology B.V. was not involved in the decision to process this data or submit this manuscript. The remaining authors declare that the research was conducted in the absence of any commercial or financial relationships that could be construed as a potential conflict of interest.

**Supplemental Materials: Detecting and measuring respiratory events in horses during exercise with a microphone: deep learning vs. standard signal processing**

Table S1: Overview of target and averaged measured GNSS speeds together with covered distance for each segment during the standardized exercise test. Note: Jog-start, Trot 1, High-speed trot and Jog-end are based on N = 15 horses while Trot 2 and Trot 3 are based on N = 13, as H1 and H2 did not perform those speeds.

|            | Target speed | GNSS speed | Distance     |
|------------|--------------|------------|--------------|
|            | km/h         | km/h       | m            |
| Jog-start  | 20           | 19.4       | 1800         |
| Trot 1     | 25           | 25.1       | 1000         |
| Trot 2     | 30           | 29.6       | 1000         |
| Trot 3     | 35           | 34.2       | 1000         |
| High-speed | 40           | 37.9       | 800* or 500** |
| Jog-end    | 20           | 18.7       | 1800         |

*: H1 and H2, **: H3 to H15.

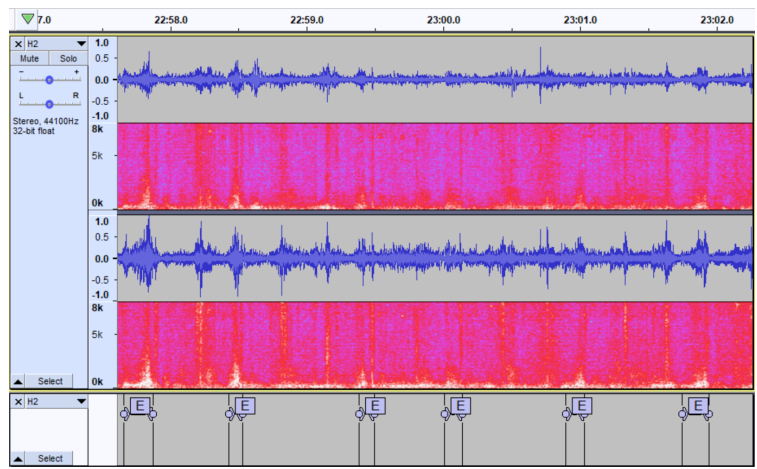

(a) Low-speed trot.

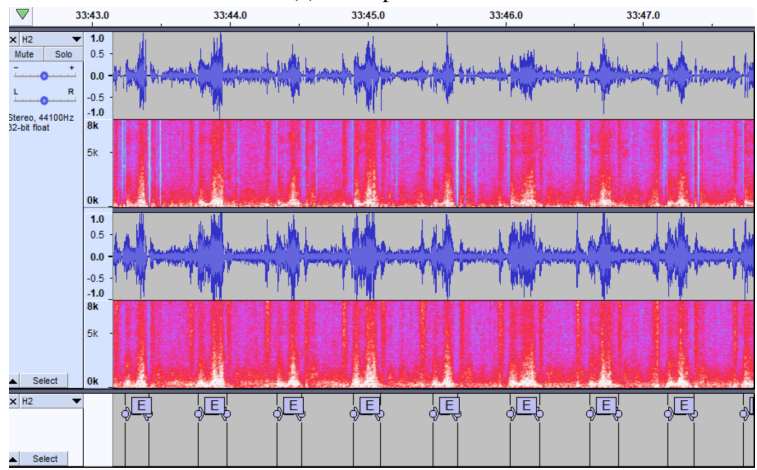

(b) High-speed trot.

Figure S1: Examples of labelled exhalation events (*E*) in the Audacity software at (a) low speed trot and (b) high-speed trot for the same horse and same duration. For each subfigure, channel 1 is the top signal and channel 2 the bottom one.

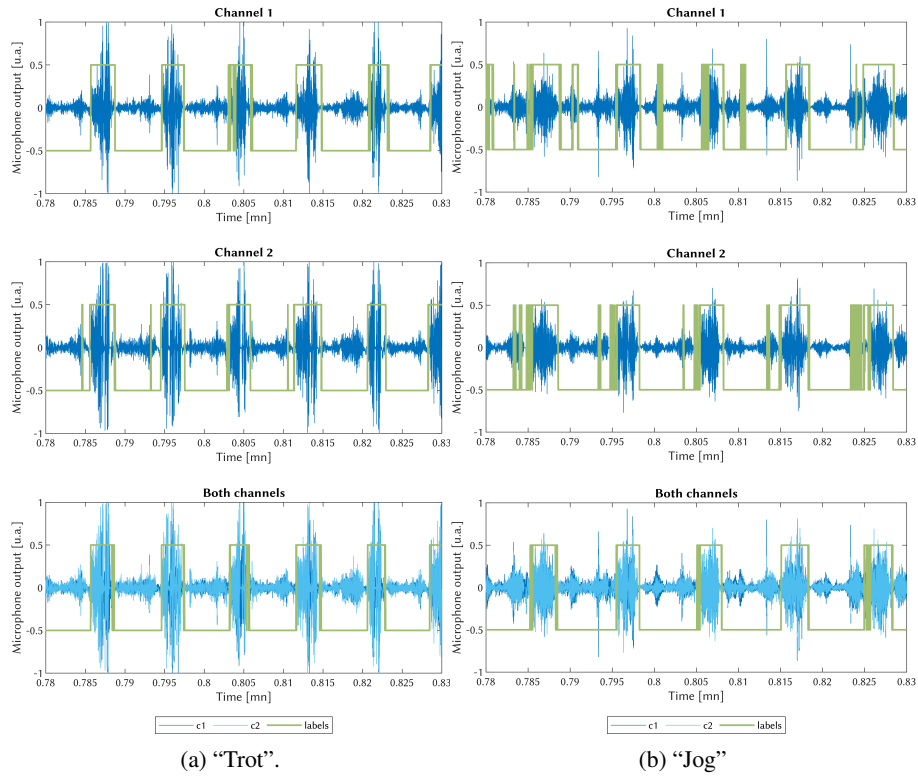

(a) "Trot".        (b) "Jog"

Figure S2: Example of obtained unprocessed exhalation events from unlabelled data and a TCN 12 model (Fs: 4410Hz) at (a) a "normal" trotting speed and (b) a "jog" trotting speed - Horse 6.

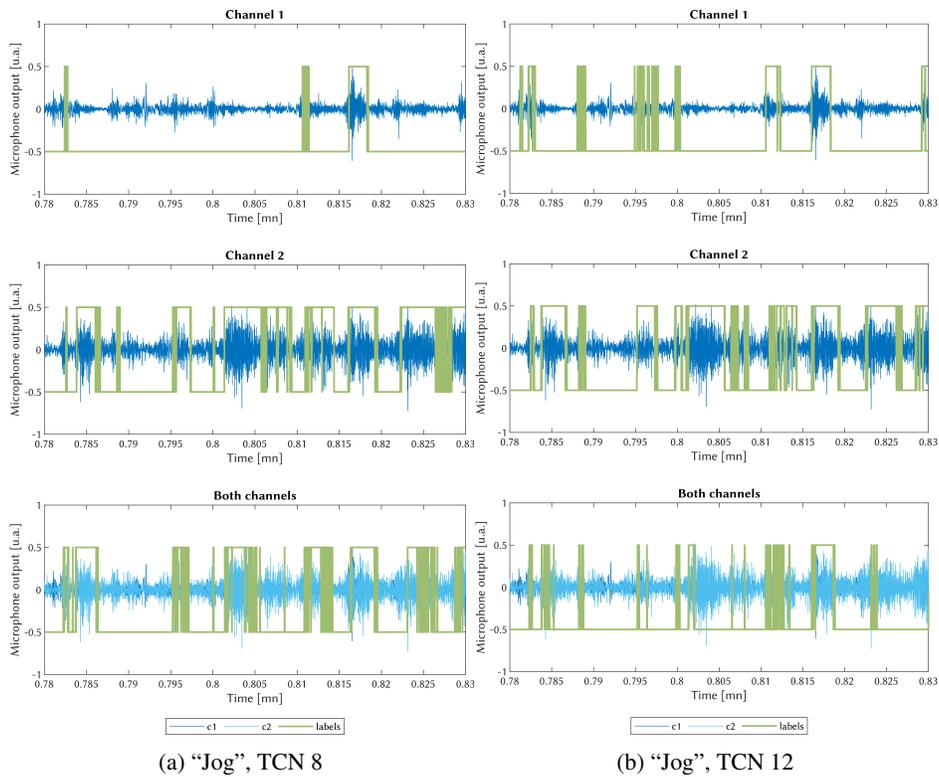

(a) "Jog", TCN 8        (b) "Jog", TCN 12

Figure S3: Example of obtained unprocessed exhalation events from unlabelled data and at "jog" trotting speed (Fs: 4410Hz) with (a) a TCN8 and (b) a TCN 12 model - Horse 3.